# Type-II Ising Pairing in Few-Layer Stanene

Joseph Falson[1], Yong Xu[2,3], Menghan Liao[2], Yunyi Zang[2], Kejing Zhu[2], Chong Wang[2], Zetao Zhang[2], Hongchao Liu[4], Wenhui Duan[2,5], Ke He[2,6], Haiwen Liu[7]*, Jurgen H. Smet[1]*, Ding Zhang[2,6]*, and Qi-Kun Xue[2,6]

[1] Max Planck Institute for Solid State Research, Stuttgart, 70569, Germany
[2] State Key Laboratory of Low-Dimensional Quantum Physics, Department of Physics, Tsinghua University, Beijing, 100084, China
[3] RIKEN Center for Emergent Matter Science (CEMS), Wako, Saitama 351-0198, Japan
[4] International Center for Quantum Materials, Peking University, Beijing 100871, China
[5] Institute for Advanced Study, Tsinghua University, Beijing 100084, China
[6] Beijing Academy of Quantum Information Sciences, Beijing 100193, China
[7] Center for Advanced Quantum Studies, Department of Physics, Beijing Normal University, Beijing 100875, China

*Correspondence to: haiwen.liu@bnu.edu.cn, j.smet@fkf.mpg.de, dingzhang@mail.tsinghua.edu.cn

**Abstract:** Spin-orbit coupling has proven indispensable in realizing topological materials and more recently Ising pairing in two-dimensional superconductors. This pairing mechanism relies on inversion symmetry breaking and sustains anomalously large in-plane polarizing magnetic fields whose upper limit is expected to diverge at low temperatures, although experimental demonstration of this has remained elusive due to the required fields. In this work, the recently discovered superconductor few-layer stanene, i.e. epitaxially strained $\alpha$-Sn, is shown to exhibit a new type of Ising pairing between carriers residing in bands with different orbital indices near the $\Gamma$-point. The bands are split as a result of spin-orbit locking without the participation of inversion symmetry breaking. The in-plane upper critical field is strongly enhanced at ultra-low temperature and reveals the sought for upturn.

Realizing superconducting materials resilient to strong external magnetic fields remains a particularly important pursuit for both applied and fundamental research alike *(1-7)*. The viability of exotic pairing mechanisms supporting this ambition is under constant scrutiny. One recent breakthrough has been the identification of "Ising pairing" in two-dimensional (2D) crystalline superconductors *(2)*. This pairing mechanism can apparently boost the in-plane upper critical field, $B_{c2,//}$, compared to the Chandrasekhar-Clogston or Pauli limit *(8-9)*. For instance, MoS$_2$ *(3-4)*, populated with charge carriers through ionic liquid gating, exhibits a $B_{c2,//}$ exceeding the value expected from the Pauli limit by a factor of 5-6. In atomically thin NbSe$_2$, $B_{c2,//}$ was reported to be as high as 31.5 Tesla — the highest applied field — at a temperature $T$ equal to 50% of the superconducting transition temperature $T_c$ *(5)*, even though the Pauli limit would only yield a field of up to 5.5 T. In amorphous superconducting films (Fig. 1**A**) spin-flip scattering, as illustrated in the cartoon in panel A of Fig. 1, has been attributed a key role in enhancing $B_{c2,//}$. However, in these high mobility crystalline samples spin-flip scattering *(10)* can be safely discarded as the origin of the enhancement as it would imply unphysically short scattering times *(3-5)*. Theory has therefore pointed to properties inherent to the band structure of these 2D materials to account for the anomalous robustness. As a result of broken inversion symmetry, opposing valleys in ***k***-space host states of opposite spin orientation (Fig. 1**B)**. By now it has been established that strong spin-orbit coupling (SOC) induces significant spin splitting among these valleys. Consequently, Cooper pairs formed from carriers in opposing valleys possess locked opposite spins and become resilient to an in-plane pair-breaking field. This physical framework inaugurated the search for ever increasing $B_{c2,//}$ almost exclusively in transition metal dichalcogenides as their crystal structure may naturally break in-plane inversion symmetry. Single layers of WS$_2$ and TaS$_2$ recently unveiled in plane critical fields that easily exceed the Pauli limit by a factor of 10 *(6-7)*.

Several theoretically prognosticated features of Ising superconductivity however remain to be verified experimentally. For example, the $B_{c2,//}$ is predicted to diverge and deviate from the 2D Ginzburg-Landau (G-L) formula at low temperatures, even if a moderate amount of disorder is present *(11-13)*. Such behavior is reminiscent of the Fulde-Ferrell-Larkin-Ovchinnikov (FFLO) state *(14-17)* (Fig. 1**C**), an epitome of robust pairing against spin polarizing fields in clean superconductors. There, macroscopic coherence gets replaced by a spatially ordered phase in the presence of a partial spin polarization at low temperatures, i.e. $T < 0.5\,T_c$. The experimental observation of a rapidly increasing $B_{c2,//}$ at low temperature provides strong support to the existence of the FFLO state in organic superconductors *(17)*. In Ising superconductors, however, it is the spin split band structure that imposes a similar renormalization to the G-L formula at $T \ll T_c$. Unfortunately, the relevant magnetic field regime in the phase diagram as $T \to 0$ is difficult to access for established Ising superconductors due to technical limitations in the attainable magnetic fields.

Here, we address this potential divergence of $B_{c2,//}$ at low temperature and breakdown of the G-L formula, that should be characteristic for Ising superconductivity, in epitaxial thin films of α-Sn(111), also referred to as few layer stanene *(18, 19)*. This material has recently emerged as a 2D superconductor. We first establish on theoretical grounds that it too is an Ising 2D superconductor although of a different type than those previously reported which derive their Ising nature from broken inversion symmetry and the spin-orbit driven splitting of valleys located at different points of the Brillouin zone. We assert that in few layer stanene, which possesses both centrosymmetry and spin-degenerate Fermi pockets at the $\Gamma$-point, spin-orbital locking and reversed effective Zeeman fields for bands with different orbital indices produce the out-of-plane spin orientation and spin splitting required for Ising pairing. We will refer to this pairing mechanism as type-II Ising superconductivity. Incidentally, in this system the in-plane magnetic fields needed to demonstrate the failure of the G-L formula and the divergence of the in-plane critical field, that so far remained elusive in experiment, are within reach.

Figure 2**A** illustrates the atomic structure of trilayer stanene grown on PbTe/Bi$_2$Te$_3$/Si(111) substrates with low-temperature molecular beam epitaxy *(18)*. Few-layer stanene itself has no M$_z$ mirror symmetry and is centrosymmetric *(20,21)*, although surface decoration and the substrate weakly break inversion symmetry of the films under study (Fig. 2**A**). Figure 2**B** sketches the band structure of the trilayer. This 3D rendering is based on angle-resolved photoemission spectroscopy (ARPES) data as well as first-principles calculations *(19)*. In the vicinity of the Fermi level, a linearly dispersing hole band surrounds a small electron pocket at the $\Gamma$-point, giving rise to two-band superconductivity *(19)*. Both the atomic and electronic structure are remarkably distinct from the widely studied transition metal dichalcogenides, for which the existence of M$_z$ mirror

symmetry, inversion symmetry breaking, and Fermi pockets at momenta that are not time-reversal invariant (near K and K') are essential for realizing Ising superconductivity. The absence of this mirror symmetry as well as broken inversion symmetry necessitates an alternative mechanism in stanene to produce the out-of-plane spin orientations required for Ising pairing. It should not rely on inversion symmetry breaking and in addition be applicable for spin-degenerate Fermi pockets near time reversal invariant momenta. Hence, we have termed this mechanism type-II Ising superconductivity *(22)* in order to distinguish it from previous instances of Ising superconductivity. We formulate our model based on ARPES results and first-principles calculations, and focus on the bands involving the $p_x$- and $p_y$-orbitals of Sn as they are the most relevant for electronic conduction. The SOC lifts the four-fold degeneracy at the $\Gamma$-point (Fig. 2 **B**) and results in two sets of spin-degenerate bands mainly composed of $(|+\uparrow\rangle, |-\downarrow\rangle)$ (solid circles in Fig. 1 **D**) and $(|+\downarrow\rangle, |-\uparrow\rangle)$ (dashed circles in Fig. 1 **D**), respectively, where $+$ and $-$ refer to the $p_x + ip_y$ and $p_x - ip_y$ orbitals *(20)*. Due to spin-orbit locking (Fig. 1**D**), bands with different orbital indices experience an opposite out-of-plane effective Zeeman field. It is parametrized as $\beta_{SO}(\mathbf{k})$ and is strongly $\mathbf{k}$-dependent. It is extraordinarily large at the $\Gamma$-point itself, where it produces a splitting of approximately 0.5 eV in monolayer stanene, which is equivalent to a field of about $10^3$ Tesla. However, it significantly weakens due to inter-orbital mixing at larger $\mathbf{k}$, since an in-plane magnetic field contributes a perturbation term to the Hamiltonian proportional to $\langle+\uparrow,-\downarrow|\sigma_x|+\uparrow,-\downarrow\rangle$, where $\sigma_x$ is the Pauli matrix. This term is zero for $\mathbf{k}=0$ and exerts increasing influence at larger $\mathbf{k}$. Even though $\beta_{SO}(\mathbf{k})$ decreases moderately with film thickness in few-layer stanene as a consequence of reduced band splitting in a quantum well setting, Ising-like pairing between $|+\uparrow\rangle$ and $|-\downarrow\rangle$ within the Fermi pockets near the $\Gamma$-point is expected to persist in few-layer stanene and this pairing is anticipated to be robust against in-plane magnetic fields.

Panel **C** in Fig. 2 presents the temperature dependent sheet resistance down to 250 mK of a sample consisting of trilayer stanene that has been grown on a 12 layer thick PbTe buffer (i.e. 3-Sn/12-PbTe). Details of the sample preparation and measurement techniques are deferred to the supplementary materials. The observed transition temperature equals 1.1 K. Figures 2 **D** and **E** display color renditions of the sample resistance in the parameter space spanned by the temperature and either the perpendicular (**D**) or the in-plane (**E**) magnetic field. They reflect the phase diagram of the superconducting ground state. The white color corresponds to approximately half of the normal state resistance and, hence, demarcates the phase transition to the normal state and also traces the temperature dependence of the upper critical magnetic fields marked by open circles. Close to the transition temperature $T_c$, both $B_{c2,\perp}$-$T$ and $B_{c2,//}$-$T$ follow the 2D G-L formula *(19)* and deviations only become apparent at lower temperatures. The out-of-plane upper critical field $B_{c2,\perp}$-$T$ exhibits an upturn which is properly captured by the formula of a two band superconductor *(23)* (solid black curve in Fig. 2**D**) that considers the orbital effect of the perpendicular magnetic field. However, when the magnetic field is applied parallel to an ultrathin superconductor, superconductivity is primarily suppressed by the paramagnetic effect and the two-band formula reduces to a simple square root dependence on $T_c$ *(23)*, indistinguishable from that of the 2D G-L formula (pink curve in Fig. 2**E**). Clearly, such a two-band treatment fails to describe the enhancement in the in-plane upper critical field observed in experiment, which amounts to 1 Tesla by cooling below $T = 0.2$ K. Moreover, the in-plane upper critical field is about one order of magnitude higher than the out-of-plane field. It exceeds the Pauli limit by a factor of 2-4 (Fig. 3 **A**), assuming the common estimate of $B_p=1.86\ T_{c,0}$. This indicates that an unusual mechanism renders the Cooper pairs robust against in-plane fields. The spin-flip scattering mechanism *(10)* can be readily ruled out, as it fails to agree with the experimental data (light blue curve in Fig. 2 **E** marked as KLB). A full theoretical derivation of the temperature dependence of $B_{c2,//}$ based on atomic orbitals of the system is presented in the supplementary materials. The lower curve in Fig. 3 **A** shows the close agreement between model and experiment. They both demonstrate a prominent upturn feature in the low temperature regime, which deviates from both the 2D G-L as well as the spin scattering formula. This upturn behavior as temperature drops is the key characteristic of Ising superconductivity that remained elusive in previous experiments. Its physical origin can be traced back to the peculiar spin split bands associated with different orbitals as shown in Fig. 1 **D,** which are protected by the crystal structure. At $T$ close to $T_c$, thermal activation results in a partially polarized system, suppressing the contribution of the spin-orbit induced spin split effect on $B_{c2,//}$. Data in this regime therefore overlap with the 2D G-L formula. As $T$ approaches zero, however, the spin orientation gets frozen and causes the upturn of $B_{c2,//}$. Quantitatively, the dimensionless parameter $\frac{\beta^*_{SO}}{T_{c,0}}$ controls the deviation point between the enhancement behavior characteristic for "Ising" superconductivity enhancement and the behavior governed by the G-L formula ($T_{c,0}$ denotes the zero field superconducting transition temperature). Note here $\beta_{SO}^*$ is the disorder renormalized SOC strength (see supplementary materials). Typically, $\beta_{SO}^*/T_{c,0} \approx 4$ in our samples (3-Sn/12-PbTe for example) and a clear up-turn appears at $T_c/T_{c,0} \approx 0.6$.

In Fig. 3 we also examine the role of the sample design to substantiate our model, in particular the number of layers of $\alpha$-Sn as well as the PbTe buffer layer thickness serve as parameters. Figure 3 **A** compares the upper critical field of two trilayer stanene samples with differing PbTe buffer layer thicknesses. The latter is known to raise the position of the Fermi level as its thickness is increased due to the donation of carriers from PbTe *(19)*. This results in a lower $T_{c,0}$ for trilayer stanene on six-layer PbTe. Although this sample possesses a higher $B_{c2,//}/B_p$ at $T \rightarrow 0$ compared to the previously examined 3-Sn/12-PbTe sample, the divergence is missing. We attribute this smoothening to the variation of the spin locking strength as one moves away from the $\Gamma$-point along the inverted Mexican hat band shape (inset to Fig. 3 **A**). Spins of the $|+\rangle$ and $|-\rangle$ orbitals are strongly locked out-of-plane at the $\Gamma$-point. This Ising-like orientation, however, becomes less favorable at larger momenta. Lowering the Fermi level therefore suppresses the spin polarization of the outer hole band, which can be simulated by an effective Rashba term in the Hamiltonian. The experimental data can be fitted well by taking into account this effect. The modified formula also nicely describes the upper critical fields of bilayer stanene (Fig. 3 **B**). Here, inversion symmetry breaking is stronger as the top Sn layer is decorated by hydrogen atoms while the bottom Sn layer sits on Te atoms of PbTe. In comparison, the middle layer of a trilayer stanene retains inversion symmetry. Following this argument, a penta-layer stanene better preserves the inversion symmetry and thus experiences weaker Rashba effect, giving rise to an apparent enhancement of $B_{c2,//}$ at low $T$ in Fig. 3 **C**. These observations highlight that the $\alpha$-Sn layer thickness is a key ingredient to reveal the delicate features of Ising superconductivity, and that there may exist a broader range of materials hosting such pairing mechanisms without the participation of inversion symmetry breaking.

14. P. Fulde, R. A. Ferrell, Superconductivity in a strong spin-exchange field. *Phys. Rev.* **135**, A550-A563.

15. A. I. Larkin, Yu. N. Ovchinnikov, Inhomogeneous state of superconductors. *Sov. Phys. JETP.* **20**, 762 (1965).

16. Y. Matsuda, H. Shimahara, Fulde-Ferrell-Larkin-Ovchinnikov state in heavy fermion superconductors. *J. Phys. Soc. Jpn.* **76**, 051005 (2007).

17. J. Wosnitza, FFLO states in layered organic superconductors. *Ann. Phys.* **530**, 1700282 (2017).

18. Y. Zang, T. Jiang, Y. Gong, Z. Guan, C Liu, M. Liao, K. Zhu, Z. Li, L. Wang, W. Li, C. Song, D. Zhang, Y. Xu, K. He, X. Ma, S.-C. Zhang, Q.-K. Xue, Realizing an epitaxial decorated stanene with an insulating bandgap. *Adv. Funct. Mater.* **28**, 1802723 (2018).

19. M. Liao, Y. Zang, Z. Guan, H. Li, Y. Gong, K. Zhu, X.-P. Hu, D. Zhang, Y. Xu, Y.-Y. Wang, K. He, X.-C. Ma, S.-C. Zhang, Q.-K. Xue, Superconductivity in few-layer stanene. *Nat. Phys.* **14**, 344-348 (2018).

20. Y. Xu, B. Yan, H.-J. Zhang, J. Wang, G. Xu, P. Tang, W. Duan, S.-C. Zhang, Large-Gap Quantum Spin Hall Insulators in Tin Films. *Phys. Rev. Lett.* **111**, 136804 (2013).

21. Y. Xu, Z. Gan, S.-C. Zhang, Enhanced Thermoelectric Performance and Anomalous Seebeck Effects in Topological Insulators, *Phys. Rev. Lett.* **112**, 226801 (2014).

22. C.Wang, B. Lian, X. Guo, J. Mao, Z. Zhang, D. Zhang, B.-L. Gu, Y. Xu, W. Duan, Type-II Ising superconductivity in two-dimensional materials with strong spin-orbit coupling. arXiv:1903.06660.

23. A. Gurevich, Limits of the upper critical field in dirty two-gap superconductors, *Physica C* **456**, 160-169 (2007).

**Acknowledgments:** We thank Yijin Zhang for fruitful discussions.

**Funding:** This work is financially supported by the National Natural Science Foundation of China (grant No. 11790311, 11604176, 51788104); the Ministry of Science and Technology of China (2017YFA0304600, 2017YFA0302902, 2018YFA0307100, 2018YFA0305603, 2016YFA0301001); and the Beijing Advanced Innovation Center for Future Chip (ICFC).

**Author contributions:** J.F. and D.Z. performed the low temperature electrical measurements with the assistance of M. L. Y. Z., K. Z., and K. H. grew the samples. Y. X., C. W., Z. Z., and W. D. carried out first-principles calculations and theoretical analysis. H.-W. L. derived the microscopic model of superconductivity with the assistance of H.-C. L. J. F., Y. X., H. L., J. H. S., and D. Z. analyzed the data and wrote the paper with input from Q.-K. X. All authors discussed the results and commented on the manuscript.

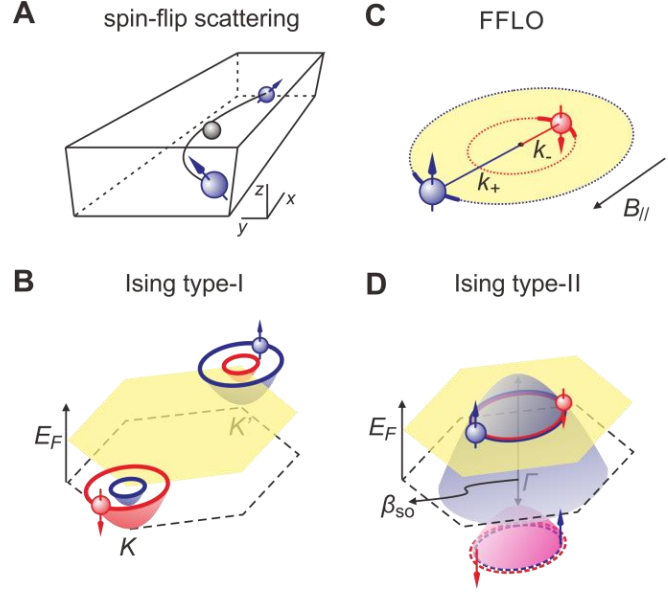

**Fig. 1. Mechanisms for an enhanced in-plane upper critical field.** (**A**) spin-flip scattering: electronic spins get randomized via scattering with impurities. (**B**) Type-I Ising superconductivity: pairing of electrons on opposite spin split valleys. Here, only one pair of electron pockets centered on K and K' points are highlighted. (**C**) Fulde-Ferrell-Larkin-Ovchinnikov (FFLO) state: Cooper pairs form with a finite momentum defined by the magnetic field. Only a small section of the Fermi surface can host pairs (solid curves). Due to this finite momentum $q$, the order parameter gets spatially modulated along the same direction, i.e. $\Delta = \Delta_0 e^{iqr}$. (**D**) Type-II Ising superconductivity: pairing of charge carriers on orbits around $\Gamma$ point with their spins aligned in the out-of-plane orientation. Hole bands are illustrated here as an example. Electron bands or bands with a more complicated dispersion are also allowed as long as the spin splitting is caused by the same SOC. The red and blue circles indicate two energetically degenerate bands with opposite spin orientations, each of which have spin split counterparts below the Fermi level (indicated by the dashed circles).

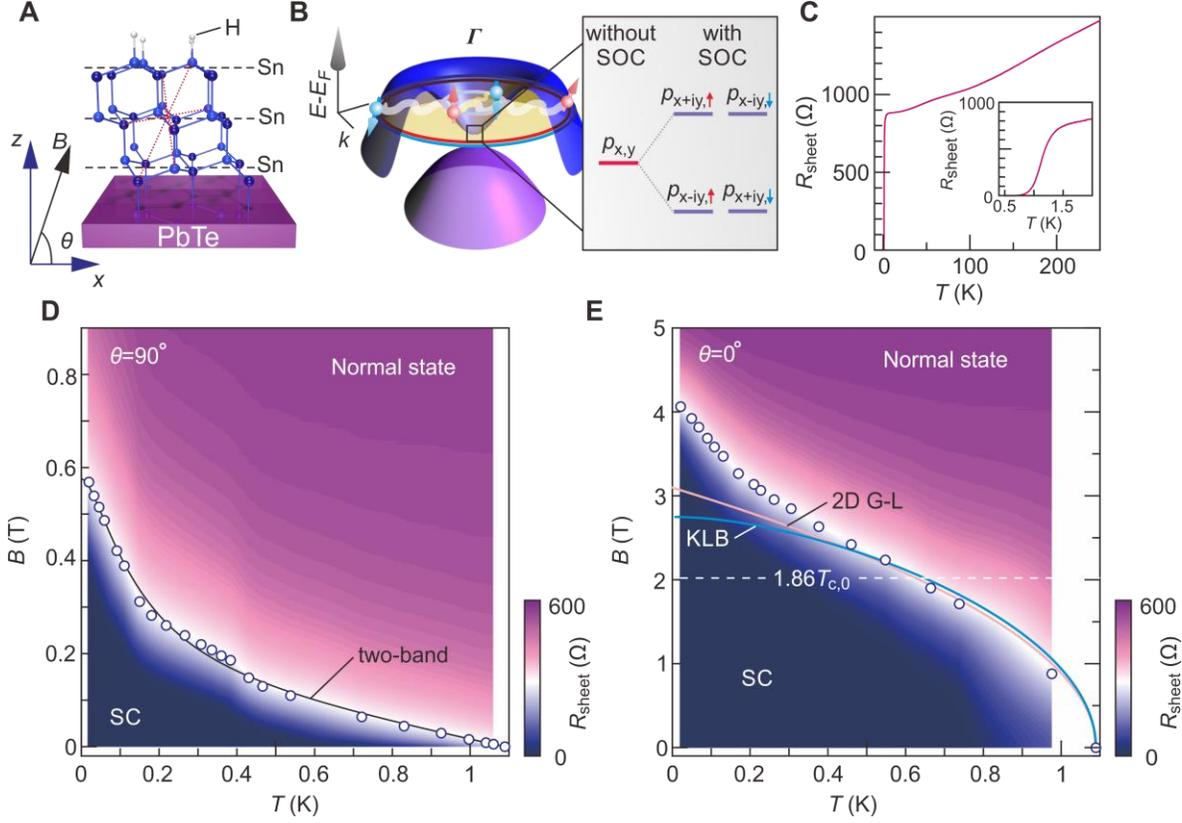

**Fig. 2. Superconducting properties of trilayer stanene.** (**A**) Atomic structure of hydrogen decorated trilayer stanene on a PbTe substrate. Dashed lines mark the three layers of Sn atoms. Red dotted lines indicate the inversion symmetry. (**B**) Three-dimensional schematic of the band structure of trilayer stanene. Blue and red circles reflect the hole/electron bands intersecting with the Fermi level. The right panel illustrates the band splitting around the $\Gamma$ point due to SOC. (**C**) Temperature dependent sheet resistance of trilayer stanene grown on 12 layers of PbTe. (**D**)/(**E**) Color coded resistance of the trilayer stanene on 12-PbTe as a function of vertical/in-plane magnetic field at a set of temperature points. The white stripe represents the boundary between the superconducting (SC) and normal state. Circles represent the magnetic fields where the resistance becomes 50% of the normal state resistance $R_n$ at a fixed $T$. Due to the smooth nature of this transition, when determining $B_{c2}$ by using another definition such as $1\%R_n$ or $10\%R_n$ would not change the general temperature dependent behavior obtained. Solid curves are theoretical fits. The solid curve in **D** is based on the formula derived for a two-band superconductor *(23)*. The blue curve in **E** is obtained using the formula that takes into account the spin-flip scattering as derived by Klemm, Luther and Beasley (KLB) *(10)*. The pink curve in **E** is based on the 2-dimensional

Ginzburg-Landau formula *(19)*. The white dashed line marks the Pauli limit using the standard BCS ratio *(3)* as well as a *g*-factor of 2.

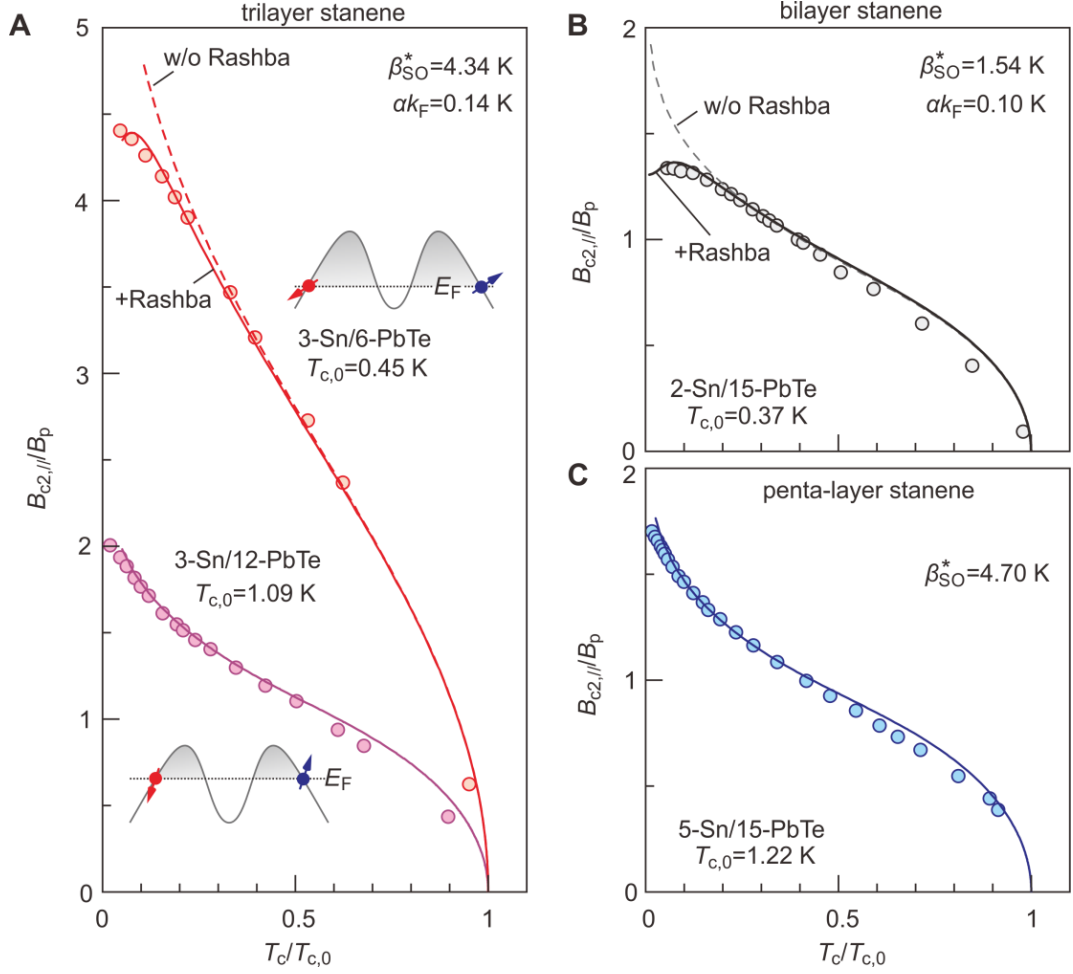

**Fig. 3. Temperature dependence of the in-plane upper critical fields in few-layer stanene samples. A** to **C** present data obtained from four samples with different stanene and PbTe substrate thicknesses. For example, 3-Sn/6-PbTe refers to a trilayer stanene grown on top of 6 layers of PbTe. The ratio of the in-plane upper critical fields, i.e. the magnetic fields at which the sample resistance becomes 50% of the normal state resistance at a given temperature, to the Pauli limit field $B_p = 1.86\, T_{c,0}$ are plotted as circular symbols. Solid/dashed curves are theoretical fits by using the formula derived for a type-II Ising superconductor (see supplementary information). The relevant fit parameters have been included for each graph.

# Supplementary information

**Methods**

Sample growth

Heterostructures were fabricated by molecular beam epitaxy *(18)*. A five quintuple layer $Bi_2Te_3$(111) buffer film was initially deposited on a Si(111) substrate, followed by an *n*-layer (n=6,12,15) PbTe(111) film. On this substrate, *n* (n=2,3,5) atomic layers of Sn were deposited at low temperature ($T \approx 150$ K) and subsequently annealed at $T \leq 400$ K to improve the surface morphology. The reflected high energy electron diffraction (RHEED) pattern displayed a streak-pattern with no detectable shift during the growth, suggesting epitaxial locking of the stanene layer with the PbTe layer. The estimated lattice constant is $a = 4.52$Å.

Low temperature measurements

All resistance data were obtained by standard low-frequency lock-in techniques. The data presented in Figure 2 **C** were collected in a $^3$He cryostat operated down to $T = 250$ mK. All other low temperature data were gathered in a top-loading dilution refrigerator with a base temperature $T \approx 20$ mK. The samples were immersed in the $^3$He-$^4$He mixture and mounted on a low-friction rotation stage to obtain both $B_\perp$ and $B_{//}$-data sets. The data shown in Fig.2 **D-E** and Fig. S1 **B** were recorded by sweeping the magnetic field from high field to zero. We estimate an error in the angle of 0.3º from the perfectly parallel field orientation. The upper critical fields are obtained from the position where the resistance drops to half of its normal state resistance (indicated by the red dotted line in Fig. S1 **B**). The data in Fig.3 **B-C** and Fig. S2 **B** were collected by sweeping the rotation angle in steps of 0.05º through the parallel configuration at a fixed total magnetic field and temperature while recording the resistance. The local minimum of each sweep forms one data point in the ($B_{//}$-$T$)-plane.

**Theory notes**

Based on previous studies *(20)*, the four-band model with the basis ($p_{x+iy,\uparrow}$, $p_{x-iy,\uparrow}$, $p_{x-iy,\downarrow}$, $p_{x+iy,\downarrow}$) and the external magnetic field *B* applied along the *x*-axis produces the following Hamiltonian:

$$H = Ak^2 + \begin{bmatrix} H_+(k) & -\mu_B B \sigma_x \\ -\mu_B B \sigma_x & H_-(k) \end{bmatrix}.$$

Here $H_\pm(k) = (M_0 - M_1 k^2)\sigma_z + v(\pm k_x \sigma_x + k_y \sigma_y)$. $\sigma_x$, $\sigma_y$ and $\sigma_z$ are the Pauli matrices. $k_x$ and $k_y$ are the momenta in the plane of the 2D superconductor. $\mu_B$ is the Bohr magneton. The energy dispersion is determined by the mass parameter $M_0$, the Fermi velocity $v$ and the material specific constants $M_1$ and $A$. We consider first the situation where only a single band crosses the Fermi level. We derive the temperature dependence of the in-plane upper critical field with the help of the Gor'kov Green function:

$$\ln \frac{T_c}{T_{c,0}} + \frac{\mu_B^2 B^2}{\beta_{so}^{*2} + \mu_B^2 B^2} Re\left[\psi\left(\frac{1}{2} + \frac{i\sqrt{\beta_{so}^{*2} + \mu_B^2 B^2}}{2\pi k_B T_c}\right) - \psi\left(\frac{1}{2}\right)\right] = 0, \text{ (SE1)}$$

where $\beta_{so}^* = \frac{\sqrt{(M_0 - M_1 k_F^2)^2 + v^2 k_F^2}}{1 + \hbar/2\pi\tau_0 \cdot k_B T_{c,0}}$ is the disorder renormalized SOC strength. $k_F$ denotes the Fermi momentum, $\psi$ the digamma function and $\tau_0$ the scattering time.

In the case where Rashba-type SOC cannot be neglected, the temperature dependence becomes:

$$\ln \frac{T_c}{T_{c,0}} + \frac{1}{2}(G_+ + G_-)\frac{\mu_B^2 B^2}{\beta_{so}^{*2} + \mu_B^2 B^2} Re\left[\psi\left(\frac{1}{2} + \frac{i\sqrt{\beta_{so}^{*2} + \mu_B^2 B^2}}{2\pi k_B T_c}\right) - \psi\left(\frac{1}{2}\right)\right] = 0, \text{ (SE2)}$$

with $G_\pm = \left(1 \mp \frac{2(\alpha k_F)^2 + \beta_{so}^{*2} - \mu_B^2 B^2}{\rho_+^2 - \rho_-^2}\right) Re\left[\psi\left(\frac{1}{2} + \frac{i\rho_\pm}{2\pi k_B T_c}\right) - \psi\left(\frac{1}{2}\right)\right]$ and $2\rho_\pm = \sqrt{(\mu_B B + \alpha k_F)^2 + (\alpha k_F)^2 + \beta_{so}^{*2}} \pm \sqrt{(\mu_B B - \alpha k_F)^2 + (\alpha k_F)^2 + \beta_{so}^{*2}}$. Here, $\alpha k_F$ denotes the disorder renormalized Rashba SOC strength.

In the more realistic situation where $M_0$, $M_1$, $A$ and $v$ give rise to the inverted Mexican hat shaped band of few-layer stanene, there are two bands crossing the Fermi level. We can join the single band formula in virtue of the quasi-classical two-band Usadel equations. The final equation that governs the temperature dependence of the critical field is

$$\frac{2w}{\lambda_0} F_1 F_2 + \left(1 + \frac{\lambda_-}{\lambda_0}\right) F_1 + \left(1 + \frac{\lambda_-}{\lambda_0}\right) F_2 = 0, \text{ (SE3)}$$

with $F_j = \ln\frac{T_c}{T_{c,0}} + \frac{1}{2}(G_{+,j} + G_{-,j})\frac{\mu_B^2 B^2}{\beta_{so,j}^{*2} + \mu_B^2 B^2} Re\left[\psi\left(\frac{1}{2} + \frac{i\sqrt{\beta_{so,j}^{*2} + \mu_B^2 B^2}}{2\pi k_B T_c}\right) - \psi\left(\frac{1}{2}\right)\right]$, $G_{\pm,j} = \left(1 \mp \frac{2(\alpha_j k_{F,j})^2 + \beta_{so,j}^{*2} - \mu_B^2 B^2}{\rho_{+,j}^2 - \rho_{-,j}^2}\right) Re\left[\psi\left(\frac{1}{2} + \frac{i\rho_{\pm,j}}{2\pi k_B T_c}\right) - \psi\left(\frac{1}{2}\right)\right]$ and $j = 1,2$. $\lambda_{11}$, $\lambda_{22}$ and $\lambda_{12}$ represent the BCS electron-phonon coupling constants, $\lambda_\pm = \lambda_{11} \pm \lambda_{22}$, $\lambda_0 = \sqrt{\lambda_-^2 + 4\lambda_{12}^2}$ and $w = \lambda_{11}\lambda_{22} - \lambda_{12}^2$.

In order to minimize the number of fit parameters, we employ equations (SE1) and (SE2) to fit our experimental data. This is valid because equation (SE3) reduces to the single band form if the two bands have similar SOC or one band dominates (neglecting interband scattering).

The extract fit parameters are summarized in Table S1. We estimate $\tau_0$ by setting the renormalization factor $1 + \hbar/2\pi\tau_0 \cdot k_B T_{c,0} = 50$, which accounts for the reduced SOC strength $\beta_{so}^*$ from the pure component originating from the electronic bands $\beta_{so}$. The estimated $\tau_0$ in few-layer stanene is comparable to that of MoS$_2$ in previous studies *(3)*.

For the perpendicular upper critical field, we use the following formula *(23)* to fit the data

$$\ln\frac{T_c}{T_{c,0}} = -\frac{\left[U\left(\frac{eD_1 B}{hT_c}\right)+U\left(\frac{eD_2 B}{hT_c}\right)+\frac{\lambda_0}{w}\right]}{2} + \left[\frac{\left(U\left(\frac{eD_1 B}{hT_c}\right)-U\left(\frac{eD_2 B}{hT_c}\right)-\frac{\lambda_-}{w}\right)^2}{4} + \frac{\lambda_{12}^2}{w^2}\right]^{\frac{1}{2}}. \quad (\text{SE4})$$

Here, $U(x) = \psi\left(\frac{1}{2}+x\right) - \psi\left(\frac{1}{2}\right)$ with $\psi$ the digamma function. $D_1$ and $D_2$ are the diffusivities of the two bands.

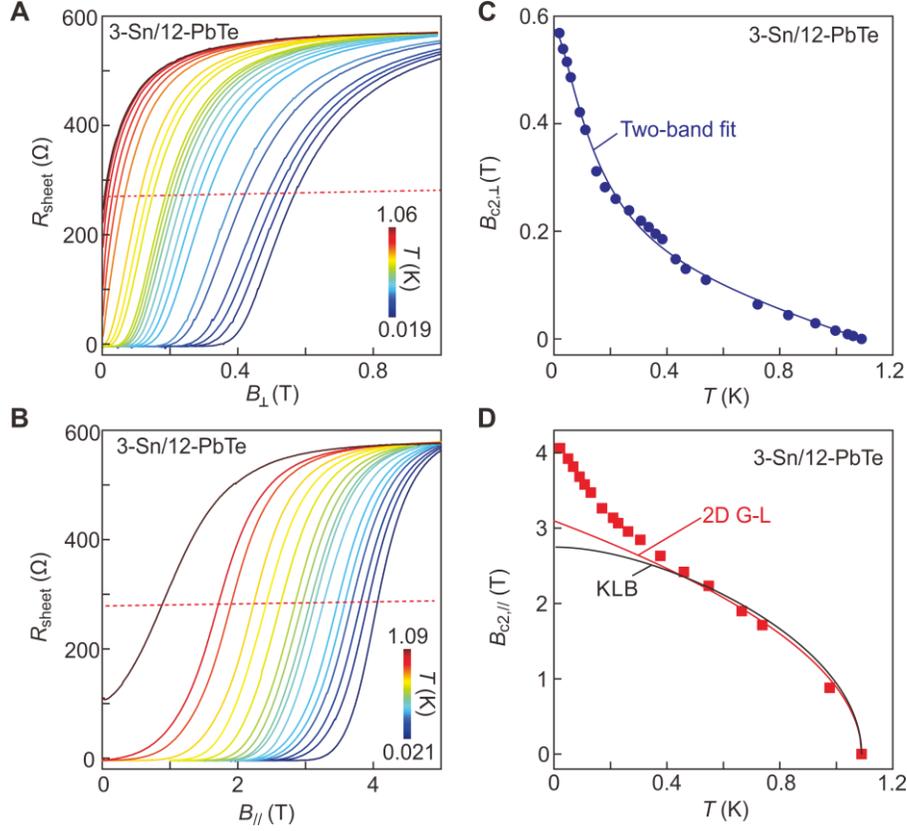

**Fig. S1: Magneto-transport data on a 3-Sn/12-PbTe sample. A, B:** Sheet resistances as a function of out-of-plane/in-plane magnetic fields at a set of temperatures. The upper critical fields are obtained from the position where the resistance drops to half of its normal state resistance (indicated by the red dotted line). **C, D:** Temperature dependence of the out-of-plane/in-plane upper critical fields. Solid curve in **C** is a fit obtained by employing the two-band formula of Eq. (SE4) *(23)*. The red and black curves in **D** are a fit using the 2D Ginzburg-Landau *(19)* and KLB formula *(10)*, respectively.

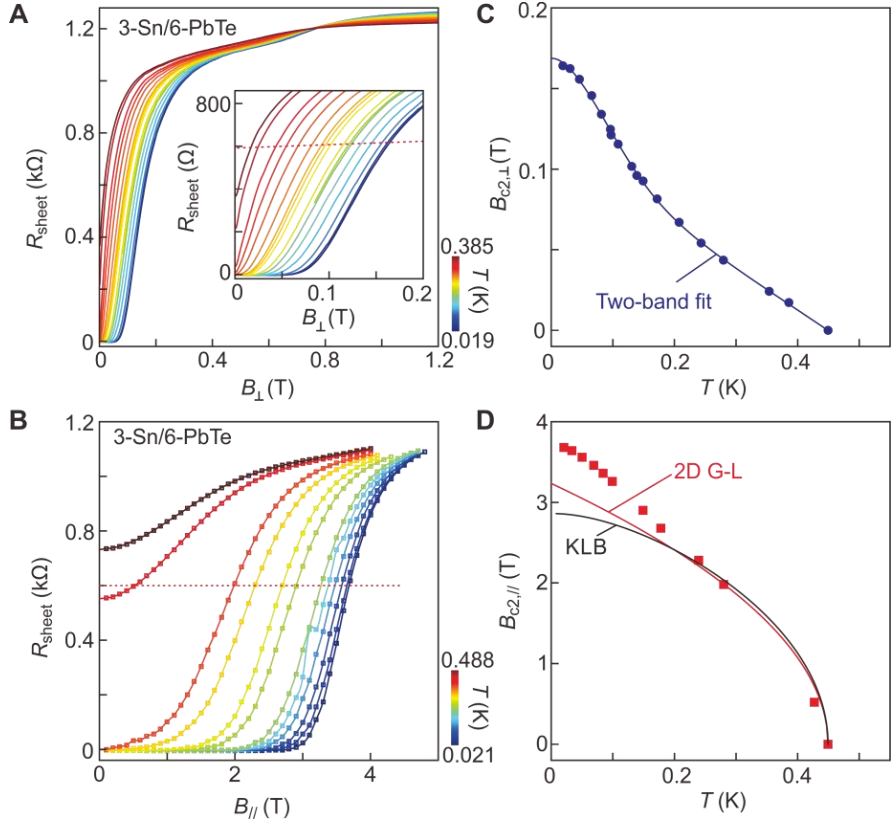

**Fig. S2. Magneto-transport data on a 3-Sn/6-PbTe sample. A, B:** Sheet resistances as a function of out-of-plane/in-plane magnetic fields at a set of temperatures. Solid curves in **B** are obtained by interpolation. The upper critical fields are determined from the position where the resistance drops to half of its normal state resistance (indicated by the red dotted line). **C, D**: Temperature dependence of the out-of-plane/in-plane upper critical fields. The solid curve in **C** is a fit employing the two-band formula of Eq. (SE4) *(23)*. The red and black curve in **D** result from the 2D Ginzburg-Landau *(19)* and KLB formula *(10)*, respectively.

|  | $T_{c,0}$ (K) | $\beta_{so}^*$ (K) | $\beta_{so}^*$ (meV) | $\beta_{so}$ (meV) | $\alpha k_F$ (K) | $\tau_0$ (ps) |
|---|---|---|---|---|---|---|
| 2-Sn/15-PbTe | 0.377 | 1.54 | 0.132 | 6.6 | 0.10 | 0.066 |
| 3-Sn/6-PbTe | 0.449 | 4.34 | 0.374 | 18.7 | 0.14 | 0.054 |
| 3-Sn/12-PbTe | 1.090 | 4.34 | 0.374 | 18.7 | -- | 0.023 |
| 5-Sn/15-PbTe | 1.219 | 4.70 | 0.405 | 20.3 | -- | 0.020 |

**Table S1. Extracted fit parameters of the four samples.** $T_{c,0}$ is the zero-field transition temperature employed in the theoretical curve. $\beta_{so}^*$ is the disorder renormalized SOC strength, which is shown in both K and meV units. $\beta_{so}$ is the corresponding intrinsic SOC strength, if assuming a renormalization factor of 50. $\alpha k_F$ denotes the disorder renormalized Rashba SOC strength. $\tau_0$ is the estimated scattering time.